# Maximum-Likelihood–Based Position Decoding of Laser-Processed Converging-Pixel CsI:Tl Detectors for High-Resolution SPECT

Xi Zhang, Arkadiusz Sitek, Lisa Blackberg, Matthew A. Kupinski, Lars R. Furenlid, Hamid Sabet

***Abstract*—This study demonstrates the feasibility of a novel fabrication technique for high–spatial-resolution CsI:Tl scintillation detectors tailored for single-photon emission computed tomography (SPECT) systems. Building upon our previously developed laser-induced optical barrier (LIOB) method—which achieved high spatial resolution, excellent sensitivity, and 100% fabrication yield in CsI:Tl detectors—we extend this approach to a converging-pixel architecture. A CsI:Tl crystal array with converging pixels was designed and fabricated, featuring entrance-face pixels of 1.6 × 1.6 mm² and photodetector-side pixels of 2 × 2 mm². To localize γ-ray interactions, both the center-of-gravity (CoG) algorithm and a maximum-likelihood (ML)–based decoding method were implemented. A custom-built four-axis motion platform was developed to deliver a finely collimated pencil beam at precisely controlled positions and angles across the array, enabling generation of a comprehensive dataset for prior knowledge and validation. The results demonstrate an energy resolution of 11.79 ± 0.53% (collimated experiment) and a position localization accuracy of 1.00 ± 0.42 mm (nearest-neighbor interpolation), confirming that the proposed converging-pixel architecture, combined with statistical decoding algorithms, provides a promising path toward the development of high-performance SPECT detectors.**

## I. Introduction

SINGLE photon emission computed tomography (SPECT) is an established nuclear medicine modality capable of capturing functional and physiological information essential for accurate disease diagnosis and therapeutic planning. Clinically, SPECT is widely used for brain function evaluation, skeletal imaging, and myocardial perfusion assessment[1], [2], [3], [4]. Despite its diagnostic value, SPECT imaging suffers from inherent limitations, particularly in spatial resolution, which remains lower than that of positron emission tomography (PET). This limitation hinders the accurate delineation of small lesions and consequently affects the precision of both diagnostic imaging and radionuclide therapy[5], [6], [7]. Conventional clinical SPECT systems based on NaI:Tl typically achieve an energy resolution of approximately 10% at 140 keV and an intrinsic spatial resolution below 5 mm[8]. As the extrinsic resolution is primarily governed by the collimator geometry (typically 6–15 mm of the full width at half maximum (FWHM) at a 10 cm distance)[9], substantial research efforts in both hardware and software development have been undertaken to emphasize high-performance detector design and dedicated scan sampling schemes, which eventually lead to improvement in the spatial resolution and sensitivity of SPECT[10], [11], [12].

Whether intended for use in PET or SPECT imaging systems, scintillator arrays are traditionally fabricated through mechanical pixelation, in which a bulk crystal is sequentially cut, polished, and separated by reflective materials to form discrete pixels. Although this technique allows for position decoding, it becomes increasingly labor-intensive and expensive as pixel sizes shrink below the millimeter scale[13], [14]. Furthermore, the inclusion of reflectors reduces the packing fraction and decreases detection sensitivity. In contrast, continuous scintillator crystals overcome these manufacturing limitations by eliminating dead space introduced by reflectors, thereby providing higher sensitivity, improved packing efficiency, and enhanced detection performance[15], [16], [17].

However, accurately decoding the interaction position of γ-photons in continuous scintillator crystals remains a major challenge, especially close to the detector edge. Positioning algorithms for such detectors can generally be classified into three categories: non-calibration methods based on light-distribution characteristics, statistical approaches relying on calibration data, and machine learning methods[18]. Among these, the center-of-gravity (CoG) algorithm is the simplest and most direct light-profile–based method, offering fast

This work did not involve human subjects or animals in its research. This work was supported in part by NIH Grant No. R01HL145160 and R01EB034785. (Corresponding author: Xi Zhang.)

X. Zhang, A. Sitek, H. Sabet are with the Department of Radiology, Massachusetts General Hospital, Harvard Medical School, Boston, MA 02129 USA (e-mail: xzhang84@mgh.harvard.edu; asitek@mgh.harvard.edu; hsabet@mgh.harvard.edu).

M. A. Kupinski and L. R. Furenlid are with the Department of Radiology and Imaging Science, and the Wyant College of Optical Sciences, University of Arizona, Tucson, AZ 85721 USA (e-mail: mkupinski@optics.arizona.edu, furen@radiology.arizona.edu).

L. Blackberg was with the Department of Radiology, Massachusetts General Hospital, Harvard Medical School, Boston, MA 02129 USA.

computation without requiring calibration data. However, it suffers from limited positioning accuracy and pronounced edge effects[19], [20], [21]. Statistical approaches, such as the Least Squares, Nearest Neighbor, and Maximum Likelihood (ML) methods, employ pre-calibrated datasets—either experimentally measured or theoretically simulated—to model the photodetector response as a function of interaction position [22], [23], [24], [25]. These methods aim to improve spatial resolution and expand the effective volume of monolithic scintillator detectors. Their primary limitation, however, lies in the dependence on accurate calibration data. Among these alternatives, the ML approach allows the incorporation of prior information and has been demonstrated to achieve superior spatial resolution. In recent years, artificial intelligence techniques have been increasingly applied in medical imaging and radiation therapy evaluation[26], [27], [28], [29].

Conventional SPECT systems typically employ either parallel-hole or pinhole collimators. Parallel-hole collimators are widely used but offer limited spatial resolution, whereas pinhole collimators provide geometric magnification at the expense of sensitivity and field of view[30]. To address this trade-off, converging collimator geometries—such as the Siemens IQ-SPECT CardioFocal collimator—have been developed, in which the collimator bores are aimed toward a focal region of interest, thereby improving both sensitivity and resolution within the targeted field of view[31], [32], [33]. More recently, the advent of detectors with high intrinsic spatial resolution has enabled a complementary strategy [34], [35], [36]: because these detectors can resolve fine spatial details inherently, image minification rather than magnification becomes feasible, allowing the detectors to be positioned in closer proximity to the patient. This compact geometry, in turn, permits a greater number of detectors to be arranged around the field of view, achieving simultaneously high spatial resolution and high sensitivity[37], [38], [39].

Given the reliability and precision of laser systems, laser-based techniques for scintillator pixelation have attracted increasing attention. In our previous work, the laser-induced optical barrier (LIOB) approach utilized tightly focused, high-intensity laser pulses to introduce localized modifications within the scintillator bulk[40], [41], [42], [43]. By accurately controlling the laser energy, pulse duration, and focusing conditions, well-defined pixel-like patterns can be inscribed without mechanical segmentation, enabling the fabrication of fine-pitch CsI:Tl scintillator arrays with 100% process yield and improved spatial resolution and detection sensitivity. Building upon this foundation, the present study extends the LIOB technique to a converging-pixel CsI:Tl crystal specifically designed for SPECT applications. Importantly, this converging-pixel geometry inherently mitigates the parallax error associated with oblique photon incidence in pinhole SPECT by aligning each pixel's longitudinal axis with the corresponding line of collimation from the pinhole aperture. As a result, gamma photons traversing a given pixel interact predominantly within that same pixel regardless of the interaction depth, effectively eliminating depth-of-interaction (DOI) ambiguity without requiring explicit DOI decoding or multi-layer detector architectures. The principle is illustrated in Fig. 1. The custom-fabricated array features entrance-face pixels measuring 1.6 × 1.6 mm² that expand to 2 × 2 mm² at the photodetector interface, guiding scintillation light toward the photosensors and enhancing photon collection efficiency. A precision four-axis motion platform (based on an earlier 5-axis prototype [44], [45]) was developed to deliver finely collimated pencil-beam radiation for algorithm calibration and validation. The resulting decoding maps were subsequently analyzed to evaluate the performance of the event-positioning algorithms. Because the conventional CoG algorithm cannot adequately resolve the pixels of such a converging-pixel, optical-barrier crystal, a maximum-likelihood (ML)–based statistical decoding method was employed; to our knowledge, this work provides the first demonstration of the complete decoding of all pixels of an LIOB converging-pixel CsI:Tl crystal.

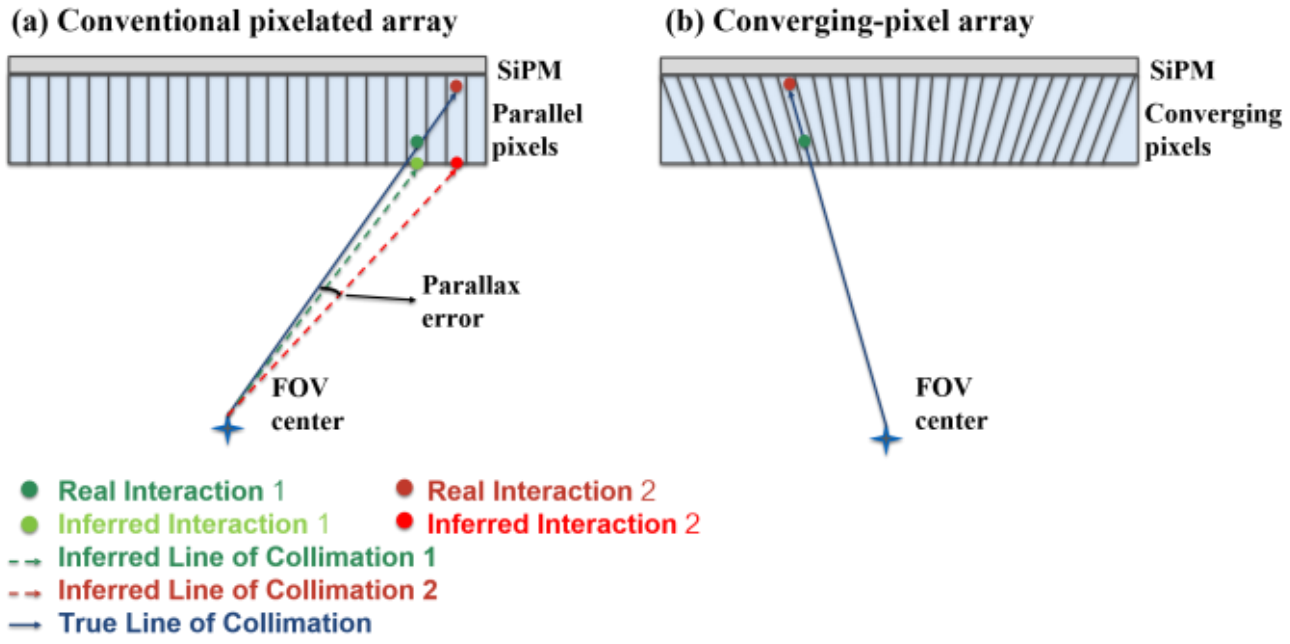

**Fig. 1.** Schematic comparison of (a) a conventional parallel-pixel array, in which the parallel pixel walls cause a DOI parallax error for obliquely incident photons, and (b) a converging-pixel array, whose walls are aligned with the lines of collimation toward the field-of-view (FOV) center so that interactions at different depths fall within the same pixel column, making the photodetector centroid depth-invariant and suppressing the parallax error.

## II. Materials and Methods

### *A. LIOB CsI:Tl Crystal*

The concept of the LIOB technique is illustrated in Fig. 2(a) and described in detail in [40]. Using this method, a CsI:Tl scintillator with dimensions of 50 × 50 × 8.05 mm³ was laser-processed into a 25 × 25 pixel array, corresponding to a pixel pitch of 2 mm at the photodetector plane and 1.6 mm at the entrance plane. Four layers of Teflon tape were wrapped around five faces of the crystal, excluding the light-output surface, to provide diffuse reflection. A photograph of the fabricated CsI:Tl crystal array is shown in Fig. 2(b). A 1.5-mm-thick light guide was optically coupled to the array using optical adhesive (DOWSIL Q2-3067) to ensure efficient optical transmission. To determine an appropriate light-guide thickness, five configurations—no light guide and light guides of 1.0, 1.5, 2.0, and 2.5 mm—were fabricated and evaluated under flood irradiation of $^{57}$Co.

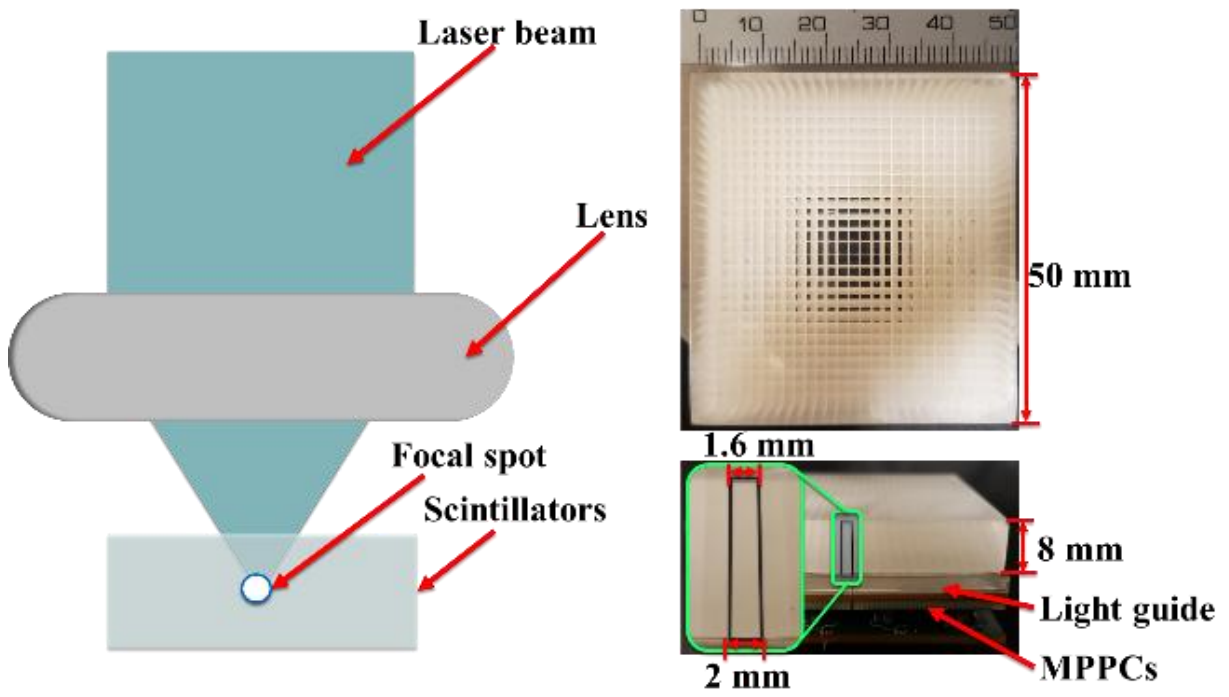

(a) (b)

**Fig. 2.** (a) Schematic illustration of the laser-induced optical barrier process; (b) photograph of the fabricated CsI:Tl crystal array.

*B. Readout Electronics*

The readout system consisted of an 8 × 8 multipixel photon counter (MPPC) array, an in-house designed 64-channel analog front-end board, two flexible printed cables (FPCs), and a CAEN V1740 waveform digitizer. A trigger threshold of 256 ADC channels was applied in the front-end electronics to suppress electronic noise. The MPPC array (model S14161-3050AS-08, Hamamatsu Photonics) comprises 64 individual pixels with a 50 μm microcell pitch and a total of 3531 microcells per pixel. The effective photosensitive area measures 25.8 × 25.8 mm², with 0.2 mm edge gaps between the active region and the package boundary. The 64-channel analog board was designed to amplify and condition the energy signals read from the MPPC array. The processed signals were then transmitted through FPCs to the CAEN V1740 digitizer for waveform acquisition. The V1740 module features a 12-bit flash ADC with a 62.5 MS/s sampling rate, suitable for mid-speed signals from inorganic scintillators coupled to MPPCs. The MPPC array was biased at 41 V using a CAEN DT5485 high-voltage power supply.

*C. Experimental setup*

To generate robust training datasets for maximum-likelihood positioning, we utilized the custom four-axis motion collimation platform (Fig. 3(a)). The system enabled precise bidirectional translation along the X and Y axes and rotational adjustment about the φ and θ angles to ensure accurate beam alignment. A tungsten collimator, designed to accommodate a capillary tube, was fabricated with a 1-mm-diameter aperture to produce a fine pencil beam. The detector, tungsten collimator, and calibrated geometry are illustrated in Fig. 3(b). The capillary was filled with a $^{99m}$Tc solution at an activity concentration of 185 MBq·cm$^{-3}$. Data were acquired over a 25 × 25 grid of collimation positions, with each position irradiated for 15 s, yielding 625 discrete datasets for algorithm training and validation. The entire experiment was completed within approximately 3 h to minimize the effect of radioactive decay on the measured activity.

For performance evaluation, a flood irradiation experiment was conducted using a $^{57}$Co point-like source with an activity of 15.4 kBq. The acquisition time was 1200 s, during which 962,990 total counts were recorded.

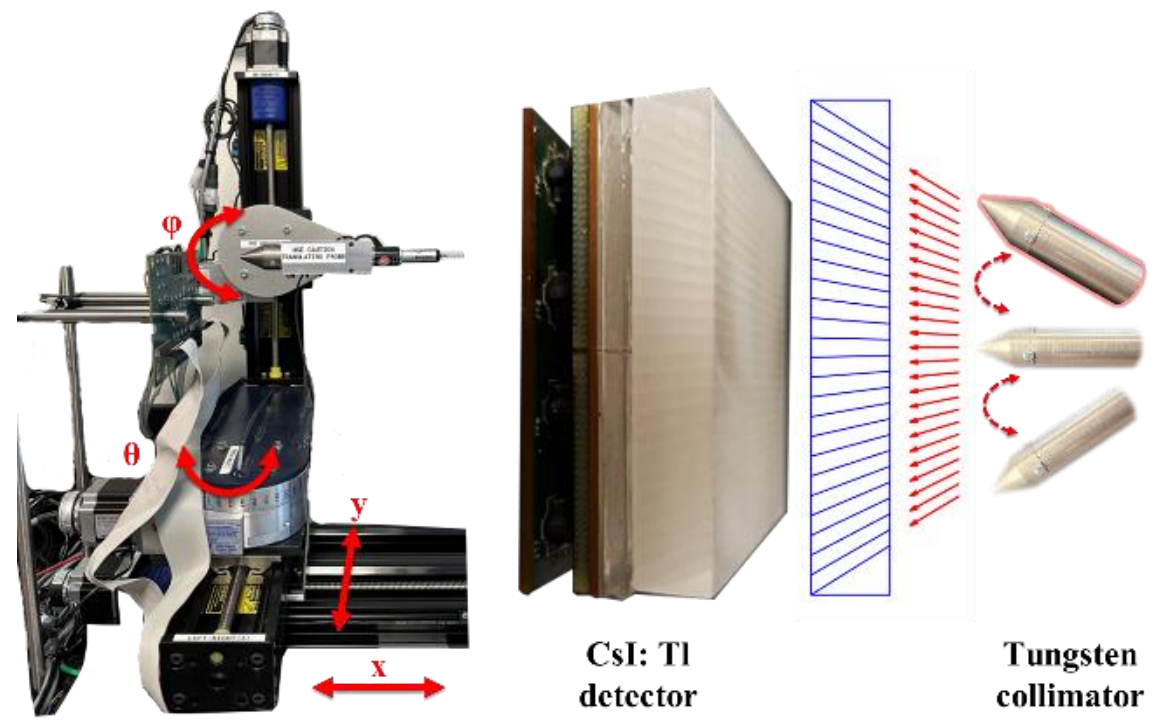

(a) (b)

**Fig. 3.** (a) Collimation system based on a four-axis motion platform; (b) CsI:Tl detector, tungsten collimator, and calibration schematic diagram.

*D. Position decoding algorithm*

*1) Center of Gravity*

The CoG method is one of the most widely used non-calibration approaches that estimates the interaction positions based on the spatial distribution of scintillation light. In this study, the photon interaction positions were determined using the CoG algorithm, as described below.

$$\begin{cases} x = \frac{P}{8} * \left(\frac{E_{r,1}+2*E_{r,2}+3*E_{r,3}+\cdots+7*E_{r,7}+8*E_{r,8}}{E_{total}} - 1\right) \\ y = \frac{P}{8} * \left(\frac{E_{c,1}+2*E_{c,2}+3*E_{c,3}+\cdots+7*E_{c,7}+8*E_{c,8}}{E_{total}} - 1\right) \end{cases} \quad (1)$$

where x and y denote the calculated interaction coordinates on a flood map of $P \times P$ pixels. $E_{r,n}$ represents the energy signals measured from the $n_{th}$ row of MPPCs. $E_{c,n}$ represents the energy signals measured from the $n_{th}$ column of MPPCs. $E_{total}$ denotes the total energy summed over all 64 MPPC channels.

*Noise suppression method:* The overall electronic noise—originating from MPPC dark current and the readout circuitry—can significantly degrade the quality of the crystal flood map. In our previous work, two simple noise-suppression strategies were introduced[46]. In this study, we employed a different approach: for each event, the 64-channel energy signals were processed by subtracting the $n_{th}$ largest signal ($N_{th}$) in each event. Negative results were set to zero, while positive values were retained as the residual differences.

$$S_{out,i} = \begin{cases} S_{in,i} - N_{th} & if \ S_{in,i} > N_{th} \\ 0 & if \ S_{in,i} \le N_{th} \end{cases} \quad (2)$$

*2) Maximum likelihood*

The ML positioning is implemented as an iterative scheme, in which the superscript (n) denotes the iteration index. The energy estimate is initialized from the total measured charge; in each iteration the log-likelihood is evaluated for every candidate pixel (Eq. (3)), the pixel that maximizes it is selected (Eq. (4)), and the energy estimate is updated from the selected pixel (Eq. (5)). These steps are repeated until the pixel assignment converges. For the case in which a γ-ray deposits energy $E$ in pixel $i$, the log-likelihood associated with the measured charge distribution $\boldsymbol{q} = \{q_j\}$ can be expressed as:

$$\mathcal{L}_i\left(i, E^{(n)} | \boldsymbol{q}\right) = \sum_j q_j \log\left[c_{i,j} E\right] - c_{i,j} E \quad (3)$$

where $q_j$ denotes the signal collected by the MPPC labeled $j$, and $\boldsymbol{q}$ represents the set of all measured charges. The coefficient $c_{i,j}$ corresponds to the fraction of scintillation photons expected to be detected by MPPC $j$ when the scintillation event originates from pixel *i,* which is also called mean detector response function (MDRF). For each pixel position, 6,000 events within an energy window of 20% (defined on the total energy) were employed to calibrate and obtain $c_{i,j}$.

The crystal pixel $i^{(n)}$ most likely to have absorbed the γ-ray is determined by the index that maximizes the log-likelihood function:

$$i^{(n)} = \underset{i}{\operatorname{argmax}}\, \mathcal{L}_i\left(i, E^{(n)}|\boldsymbol{q}\right) \quad (4)$$

where the superscript *(n)* refers to the iteration number. Likewise, $E^{(n)}$ denotes the energy estimate after the $n_{th}$ update. This result can be employed to refine the energy estimate after the pixel with the highest likelihood is identified.

$$E^{(n+1)} = \frac{\sum_j q_j}{\sum_j c_{i^{(n)},j}} \quad (5)$$

In practice, the MDRF is typically sampled on a coarse spatial grid and then interpolated to obtain values on a finer grid. In this study, three interpolation schemes—bilinear, bicubic, and nearest-neighbor—were applied to the MDRF. Specifically, the MDRF, of size 64 × 625 (64 MPPC channels × 625 calibration positions), was interpolated to 64 × (256 × 256) and 64 × (512 × 512). Nearest-neighbor interpolation assigns to each fine-grid point the value of the nearest calibrated position, producing a piecewise-constant response that preserves the sharp pixel boundaries; bilinear interpolation estimates each value as a distance-weighted linear combination of the four nearest calibrated samples; and bicubic interpolation uses the surrounding 4 × 4 neighborhood (16 samples) to produce a smoother, continuous response. The three interpolated matrices were then used in the ML decoding to compare and optimize the decoding performance.

### *E. Assessment of the energy performance and decoding map*

*Energy resolution:* The energy resolution was determined from the energy spectrum, which was obtained by summing the signals of all 64 MPPC channels for each event. Following the NEMA NU-1 standard, the photopeak was fitted with a Gaussian function, and the energy resolution was calculated as FWHM of the photopeak divided by its centroid energy, expressed as a percentage. For the collimated dataset, the energy resolution was evaluated for each of the 625 crystal pixels and averaged, whereas for the flood-irradiation dataset it was obtained from the global energy spectrum.

*Peak-to-valley ratio*: The peak-to-valley ratio (PVR) is one of the primary metrics for assessing the decoding performance of PET and SPECT detectors. In this study, the PVR values were derived from the line profiles obtained by summing the decode maps along each direction, computed from the central 23 crystal peaks, and then averaged over the row and column directions. To avoid evaluating the decoding algorithms on data that contributed to the MDRF, the collimated events acquired at each position were randomly partitioned into two non-overlapping subsets of 6,000 events each: the first subset was used to calibrate the MDRF coefficients $c_{i,j}$ in Section II.D.2, while the second, independent subset was held out exclusively for the decoding performance evaluation reported throughout this section. For the collimated dataset, a total of 6000 × 25 × 25 events were aggregated, decoded based on the MDRF, and used to generate histograms for PVR evaluation. For the flood-irradiation dataset, the flood maps were directly decoded using the ML algorithm, followed by PVR analysis.

*Segmentation and position error*: To quantify the decoding accuracy, regional segmentation of the decoded maps was performed. The overall workflow of the decoding-error calculation is summarized in Fig. 4. Different interpolation methods employed distinct segmentation strategies. In detail, the overall decoding map was projected onto the horizontal and vertical axes by summing the event counts along its columns and rows, producing two one-dimensional count profiles. In each profile, the 25 peaks corresponding to the 25 columns (or rows) of crystals were located using a peak-detection procedure with a prescribed minimum peak-to-peak separation, and the boundary between two neighboring crystals was defined as the midpoint between their adjacent peaks, with the two outer edges of the map taken as the remaining boundaries. The resulting 26 boundaries along each axis form a 25 × 25 grid that serves as a look-up-table, assigning every decoded event to a specific crystal pixel. For the bilinear and bicubic interpolations the peaks were located on the smooth, continuous decoding maps, whereas the nearest-neighbor method produced discrete decoded points, which made the segmentation straightforward and unambiguous. For the CoG method, however, the peripheral crystals do not produce distinct peaks in the profiles owing to the pronounced edge effect; for these edge regions the boundaries were therefore obtained by partitioning the peripheral map uniformly into the corresponding number of crystals, while the central crystals were still segmented by the peak-midpoint method. After segmenting the flood maps, decoding errors were evaluated. Specifically, because the true crystal location (x, y) of each collimation position is known, the events acquired at that position were assigned to crystal pixels (i, j) using the segmentation look-up-table, and the positional error of the position was computed as the count-weighted mean distance between the assigned pixels and the true location: the number of events assigned to each pixel was multiplied by the Euclidean distance from that pixel to the true crystal position, summed over all pixels, and divided by the total number of events. This quantity was evaluated both for the 25 collimation positions along the central column and for all 625 collimation positions, and the overall positional error is reported as the mean ± standard deviation over all positions.

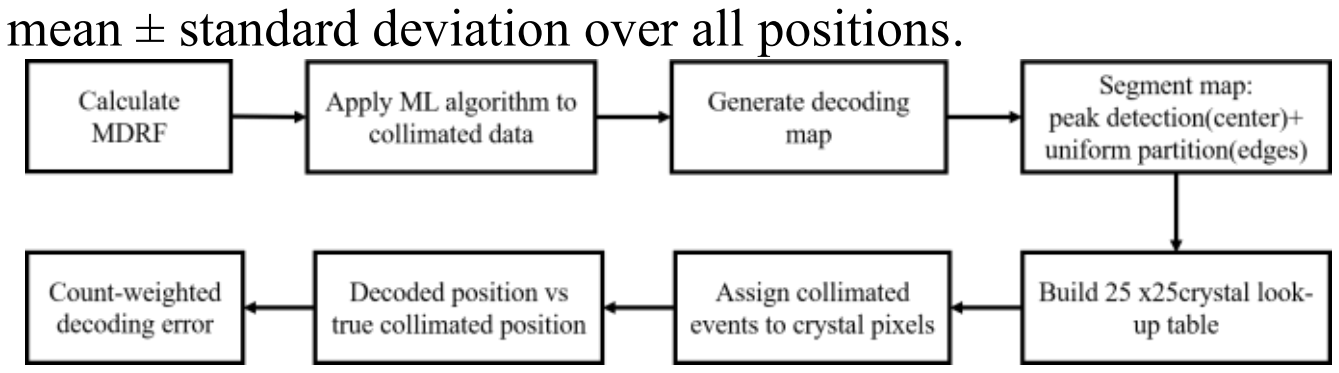

**Fig. 4.** Workflow of decoding error calculation.

*Correct- and adjacent-crystal decoding fractions*: As an additional measure of decoding accuracy that is independent of whether the output is discrete or continuous, the fraction of counts decoded to the correct crystal pixel and to its immediate neighborhood was computed from the collimated data. For each collimation position, whose true crystal location is known, the decoded events were assigned to crystal pixels using the segmentation look-up table, and the cumulative fraction of counts falling within k crystals (Chebyshev distance) of the true crystal was evaluated: k = 0 corresponds to counts decoded to the correct crystal, and k ≤ 1 to counts decoded within ±1 crystal, i.e., within the 3 × 3 neighborhood centered on the true crystal. The overall fractions were obtained as count-weighted averages over all 625 collimation positions, and separate averages were computed for the central 5 × 5 region and the outermost five-crystal-wide ring. It should be noted that the correct-crystal fraction is a conservative lower bound of the intrinsic decoding accuracy: the 1-mm collimator aperture is comparable to the 1.6-mm pitch of the crystal entrance face, so the finite beam spot, beam divergence and penetration, and inter-crystal Compton scatter place a nonnegligible fraction of the true interactions in crystals immediately adjacent to the nominal beam position. The within-±1-crystal fraction is therefore reported alongside, as it separates genuine decoding failures from these physical effects. In addition, 25 × 25 maps of the per-position correct-crystal and within-±1 fractions were generated for each method to visualize the spatial distribution of the decoding accuracy.

## III. Results

### A. Energy resolution

The energy resolution measured from the flood irradiation experiment was 15.76% at 122 keV ($^{57}$Co), and the corresponding energy spectrum is shown in Fig. 5(a). From the collimated measurements acquired at 625 positions, an individual energy spectrum was obtained, yielding an average energy resolution of 11.79 ± 0.53% at 140 keV ($^{99m}$Tc). The spectra corresponding to the pixel located at the corner, and the central pixel (highlighted in red in Fig. 5(b)) are shown in Fig. 5(c) and Fig. 5(d), respectively. The photopeak position and energy resolution were measured for all five light-guide configurations under identical $^{57}$Co flood irradiation. The measured energy resolution was 16.43%, 16.15%, 15.76%, 16.02%, and 16.48% for the no-light-guide, 1.0, 1.5, 2.0, and 2.5 mm configurations, respectively. The measured photopeak was 25619, 24878, 24351, 23625, and 23242 ADC channels for the no-light-guide, 1.0, 1.5, 2.0, and 2.5 mm configurations, respectively. The energy resolution remained within 15.76–16.48% and the photopeaks remained within 23242-25619 ADC channels across all configurations, indicating that the energy performance of the detector is weakly sensitive to the light-guide thickness over the 0–2.5 mm range. The 1.5 mm light guide was ultimately selected, as it provided the best energy resolution while maintaining a relatively high photopeak position.

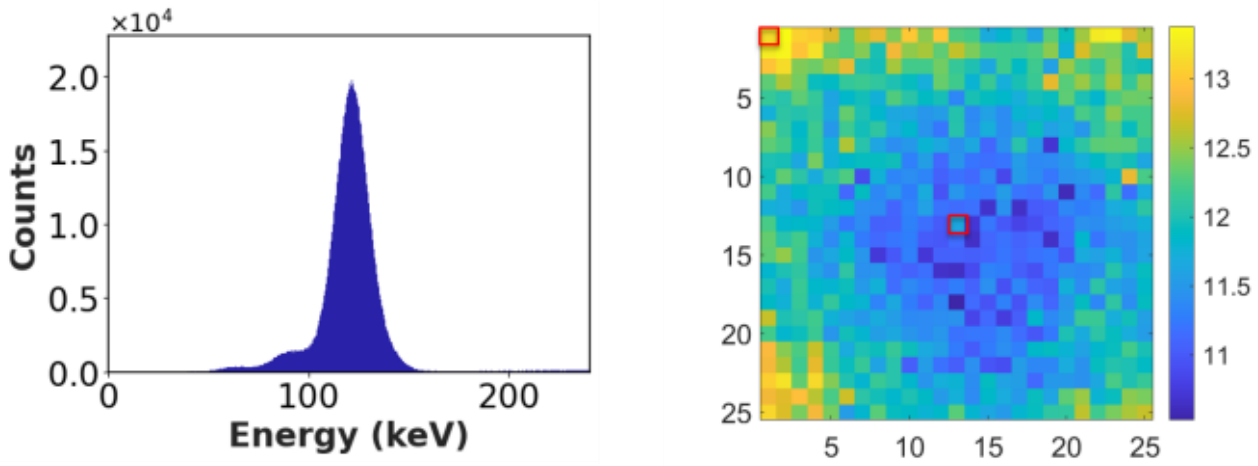

(a) (b)

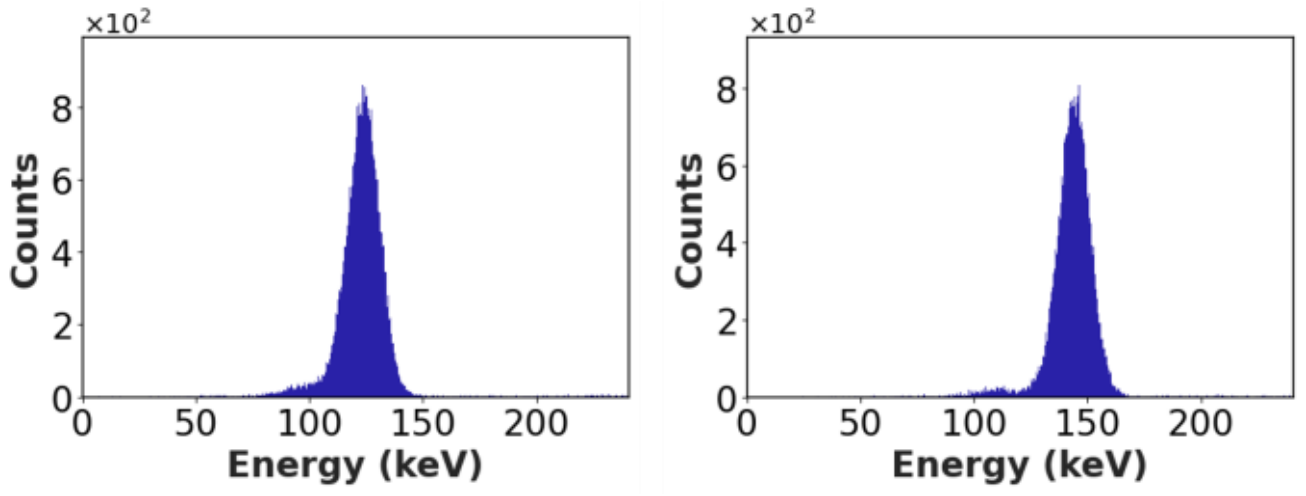

(c) (d)

**Fig. 5.** (a) Energy spectrum obtained from the flood irradiation experiment; (b) energy resolution map derived from the calibrated 625 collimation positions, with two selected crystal pixels marked by red squares; (c) energy spectrum from the pixel located at the corner; and (d) energy spectrum from the central pixel.

### B. CoG Results

#### 1) Flood irradiation data

The CoG method was applied to decode the flood irradiation data within an energy window of 20%, and the corresponding results are shown in Fig. 6. Fig. 6(a) and 6(b) present the decoding maps obtained without and with the noise-suppression method, respectively. Parameter *n* in equation (2) was optimized and set to 16. Fig. 6(c) and 6(d) show the horizontal event-count profiles (projected across all columns) along the x-direction. Approximately 15 × 15 pixels in the central region could be clearly identified. However, the pixels at the edges and corners were strongly merged, making them indistinguishable. In addition, the nonuniform distribution of the decoded spots complicated map segmentation for subsequent image reconstruction.

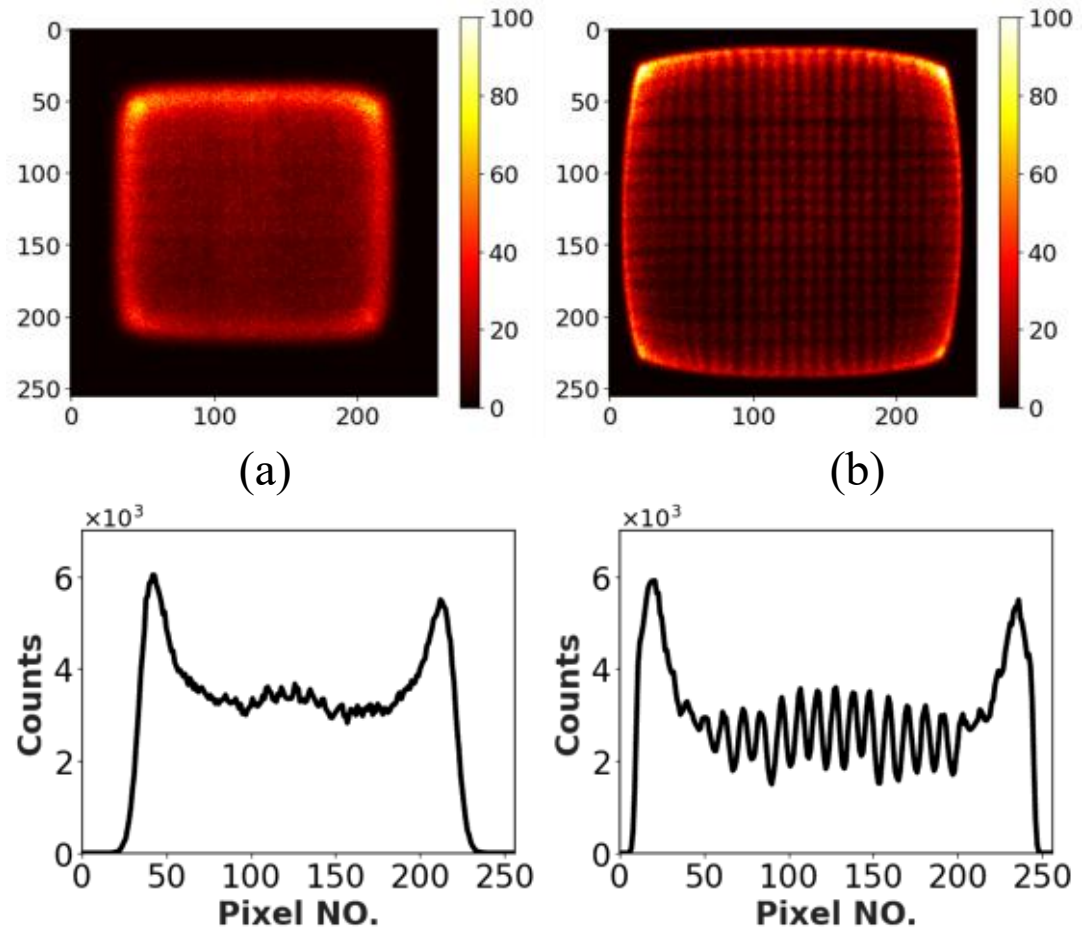

(c) (d)

**Fig. 6.** Decoding maps of the flood irradiation results obtained (a) without and (b) with the noise-suppression technique; event-count profiles (c) without and (d) with noise suppression. The profiles in (c) and (d) are projections of the decoding map across all columns.

#### 2) Collimated data

A total of 625 datasets were acquired in the collimation experiments. For each collimation position, 6000 events within a 20% energy window (defined on the total energy, i.e., the summed signal of all 64 MPPC channels) were randomly selected and combined to construct the complete collimated dataset. The data quality was preliminarily evaluated using the CoG method. Fig. 7 shows the decoding maps and the corresponding profile plots. Because the calibration data covered only a portion of each pixel, the decoded spots in the maps appeared clearer than those from the flood irradiation data. However, overlapping among pixels remained noticeable at the edges and corners.

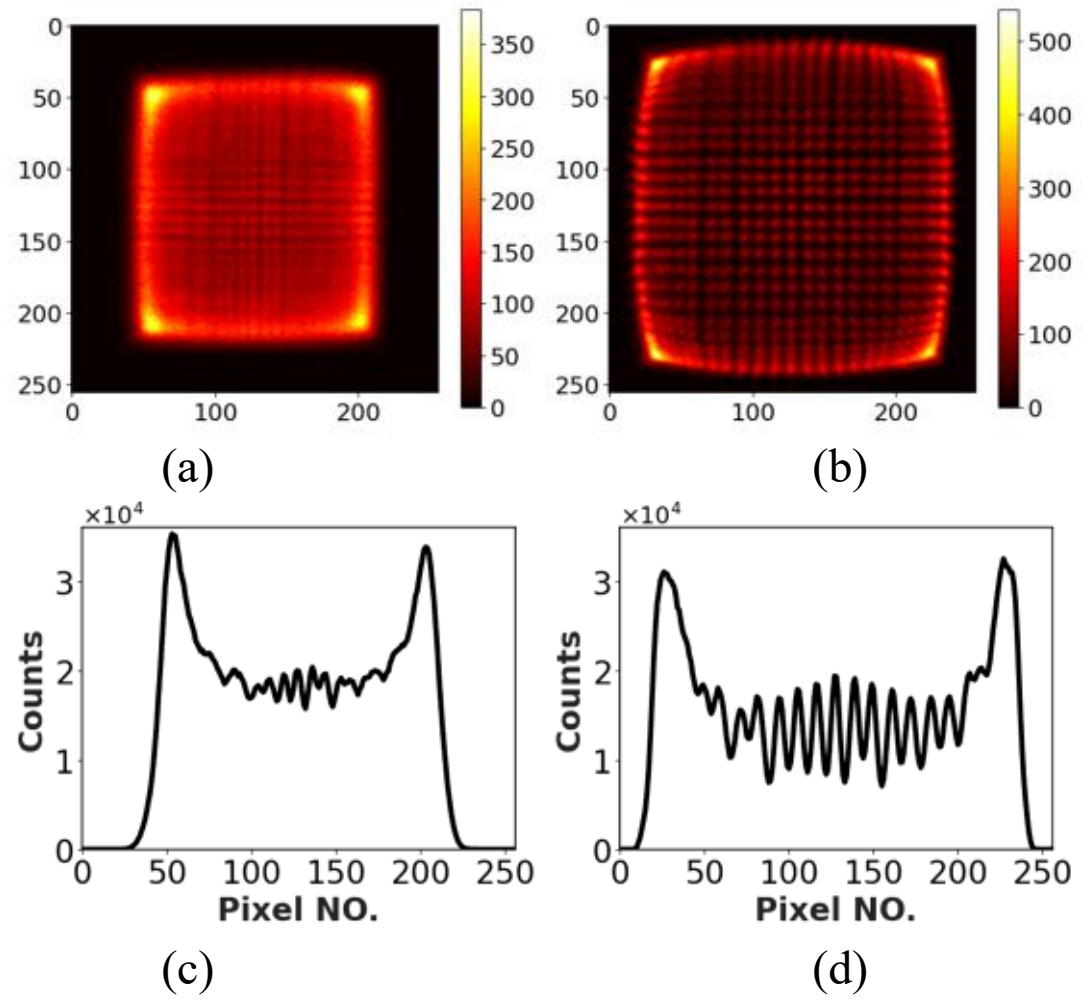

**Fig. 7.** Decoding maps of the collimated results obtained (a) without and (b) with the noise-suppression technique using the CoG algorithm; event-count profiles (c) without and (d) with noise suppression. The profiles in (c) and (d) are projections of the decoding map across all columns.

Decoding error analysis was also carried out for the CoG method. Because the CoG decoding map does not exhibit distinct peaks for the peripheral crystals, the central crystals were segmented using the peak-midpoint method and the edge regions by uniform partition, and a 25 × 25 look-up-table was established to determine the crystal pixel to which each event was decoded. The decoded crystal positions were then compared with the actual collimation positions. As shown in Fig. 8, 25 collimation datasets along the central column were analyzed (Fig. 8(a) and (b)), and the decoding error increased from the central pixels toward the edges; across all 25 × 25 pixels the error averaged 1.44 ± 1.26 mm, with a minimum of 0.21 mm (Fig. 8(d)). Using the collimated data with known beam positions, the fraction of counts decoded to the correct crystal was 52.1% overall (81.0% within the central 5 × 5 region and 38.8% in the outer five-crystal ring), and the fraction decoded within ±1 crystal of the true position was 90.3% overall (98.0% central, 86.1% outer ring) (Fig. 16(a)).

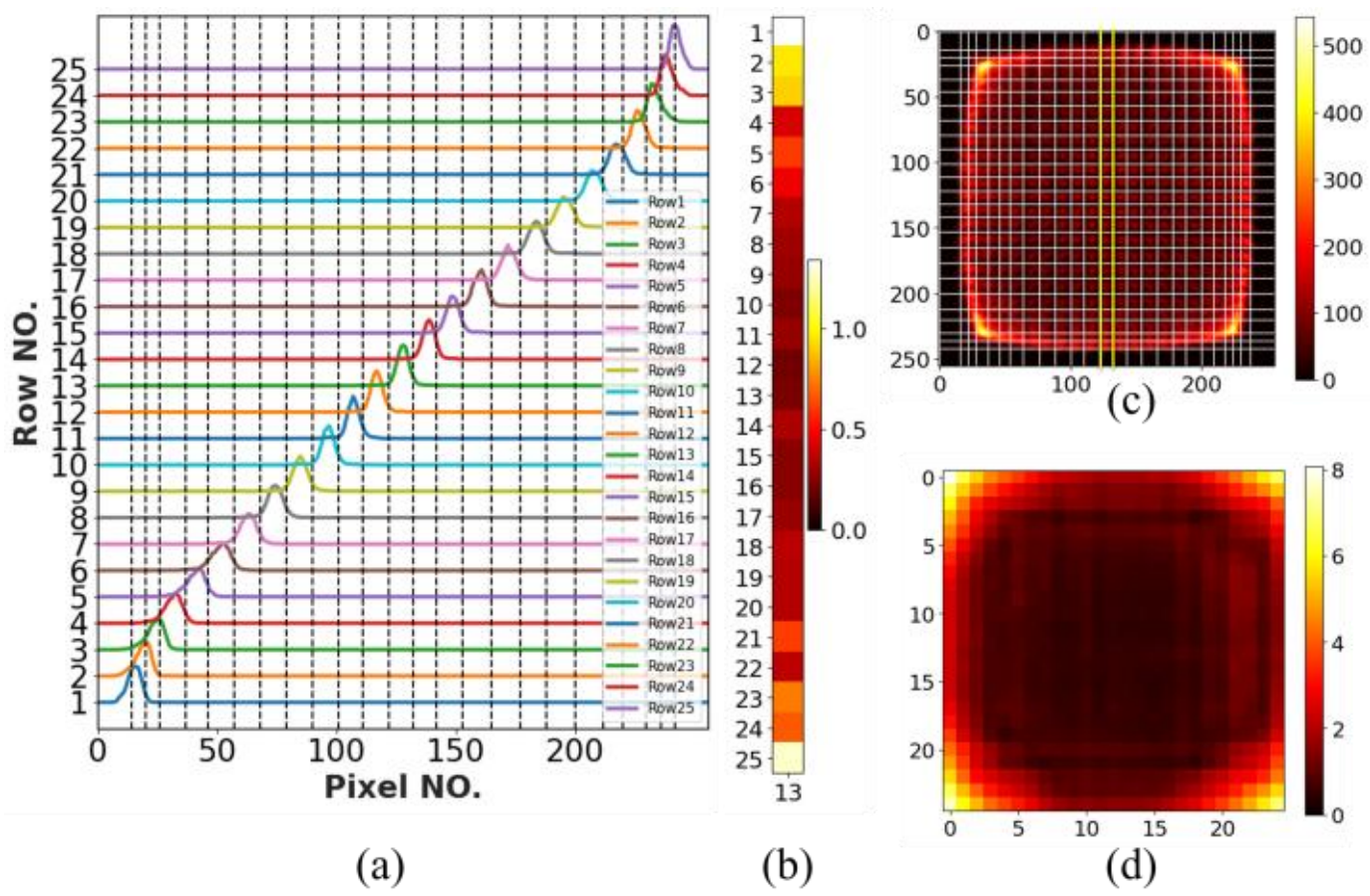

**Fig. 8.** CoG method: (a) event-count profiles of the 25 collimation datasets along the central column; the black dashed lines indicate the image segmentation boundaries; (b) positional errors of the selected datasets; (c) segmented decoding map, where the decoding region corresponding to the selected 25 datasets is highlighted between the two yellow lines; (d) positional errors of all 25 × 25 collimation datasets.

## C. ML results

### 1) Measured MDRF

Based on the datasets from the collimated experiments, the coefficient $c_{i,j}$, representing the fraction of scintillation photons expected to be detected by MPPC *j* when the scintillation event originates from pixel *i,* was calculated and shown in Fig. 9(a). Because the coefficient values were very close to each other, even minor variations in the likelihood could decisively affect the event assignment, occasionally resulting in a one-pixel misplacement. Fig. 9(b)–9(d) display the interpolated MDRFs with a size of 64 × (256 × 256), illustrating that the differences among the interpolation methods were minimal.

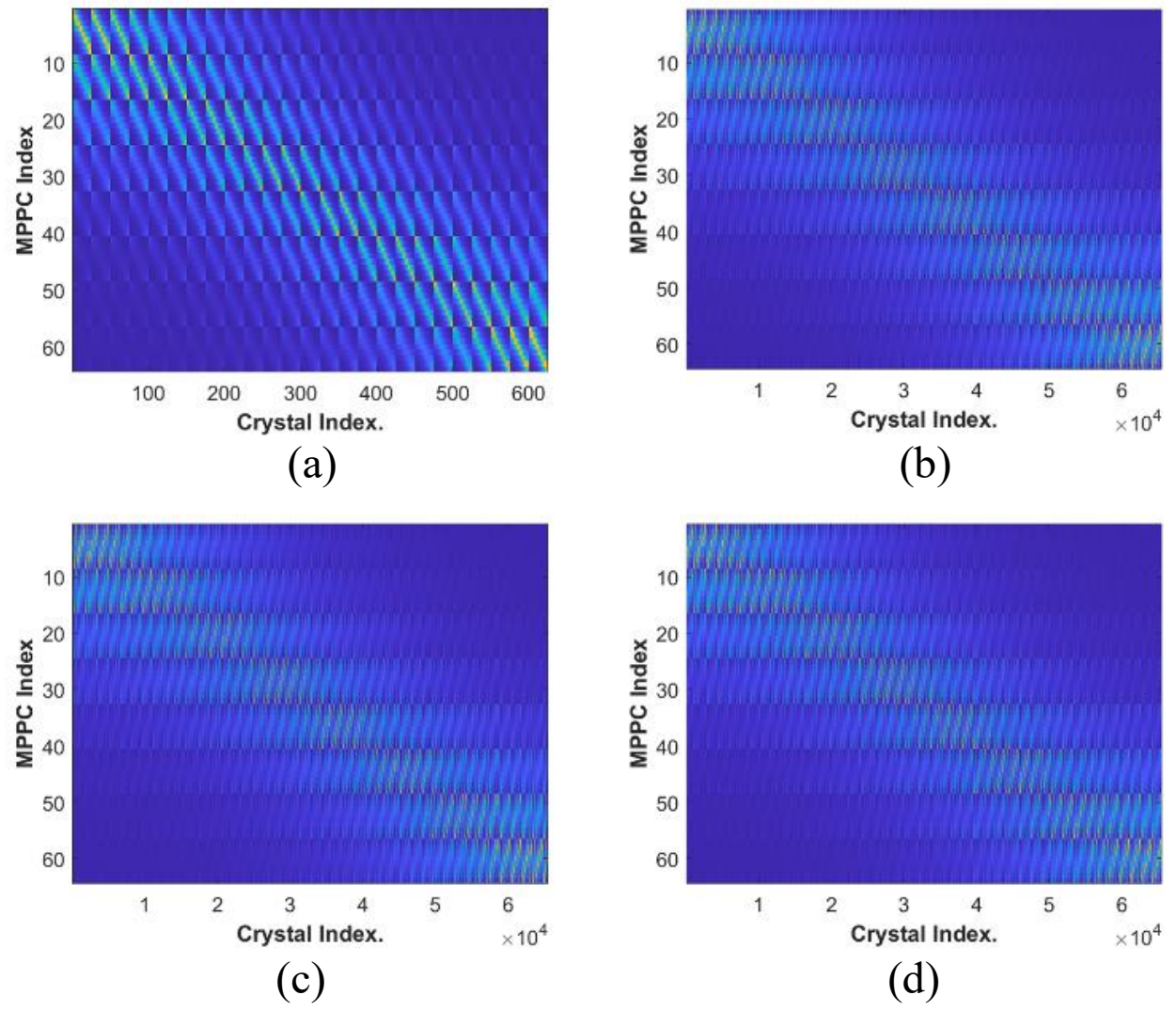

**Fig. 9.** (a) Scintillation light-response matrix (64 × 625); interpolated light-response matrices obtained using (b) bilinear, (c) bicubic, and (d) nearest-neighbor interpolation methods (64 × 256 × 256).

### 2) Collimated data

*Bilinear method:* Fig. 10 shows the decoding maps and count-profile plots obtained from the collimated dataset using bilinear interpolation. The mean PVRs, averaged over the row and column directions, were 2.28 and 2.94 for pixel resolutions of 256 and 512, respectively. The decoding maps indicate that the datasets corresponding to different collimation positions were relatively uniformly distributed across the image, without exhibiting the compression typically observed in the CoG decoding maps. However, in the profile plots, only about 13 peaks and valleys were distinguishable, and the decoding map clearly resolved only an approximately 8 × 8 crystal region. The decoding performance in the edge and corner regions was, to some extent, inferior to that of the CoG method. In addition, a considerable number of artifacts were observed in the decoding maps, including abnormal patterns on the decoded spots and unusually bright regions along the edges.

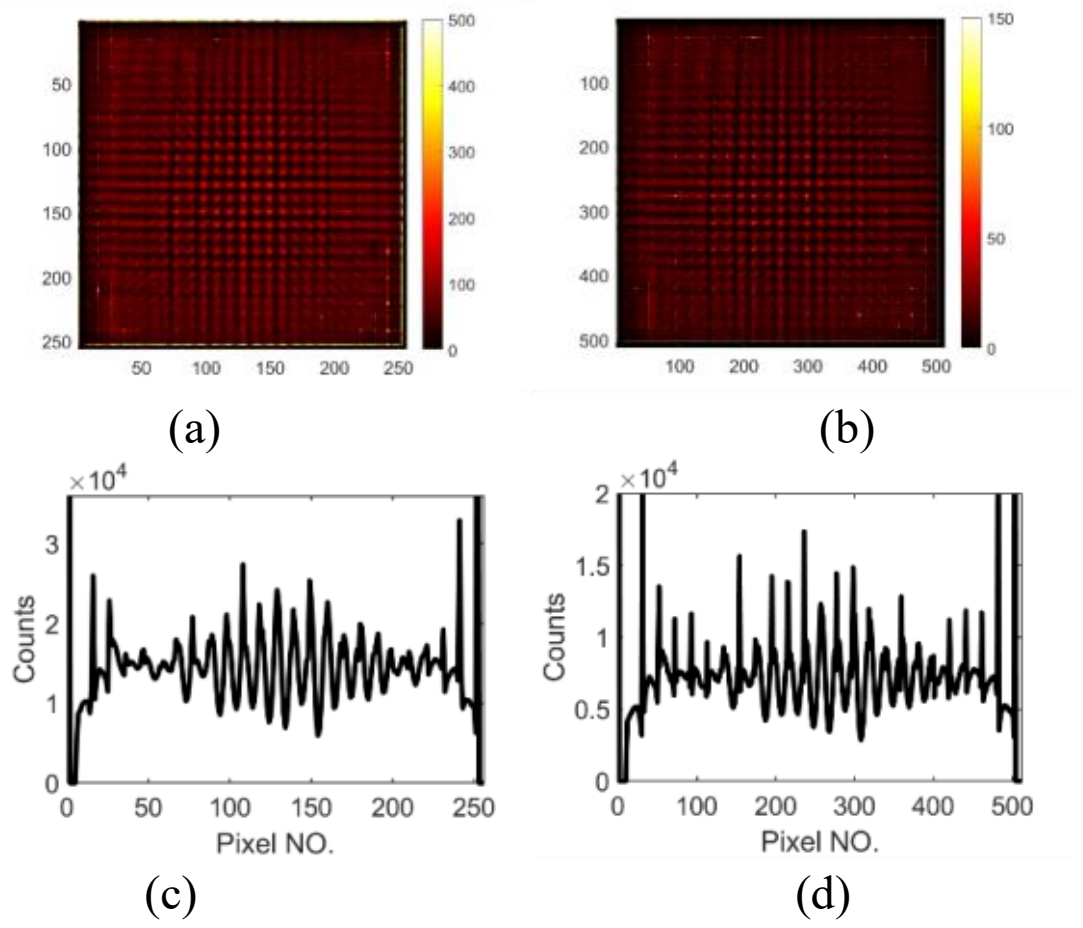

**Fig. 10.** Decoding maps of the collimated experiments obtained using bilinear interpolation with the ML algorithm at pixel resolutions of (a) 256 × 256 and (b) 512 × 512; corresponding event-count profiles are shown in (c) and (d), respectively.

For the bilinear interpolation, the 25 collimation datasets along the central column were analyzed (Fig. 11(a) and (b)); across all 25 × 25 pixels the average positional error was 1.14 ± 0.48 mm (Fig. 11(d)), with a minimum of 0.23 mm. The same central-to-edge trend remains. The corresponding fraction of counts decoded to the correct crystal was 57.3% overall (79.2% central 5 × 5, 49.2% outer ring), and 96.3% of the counts were decoded within ±1 crystal (98.0% central, 95.5% outer ring) (Fig. 16(b)).

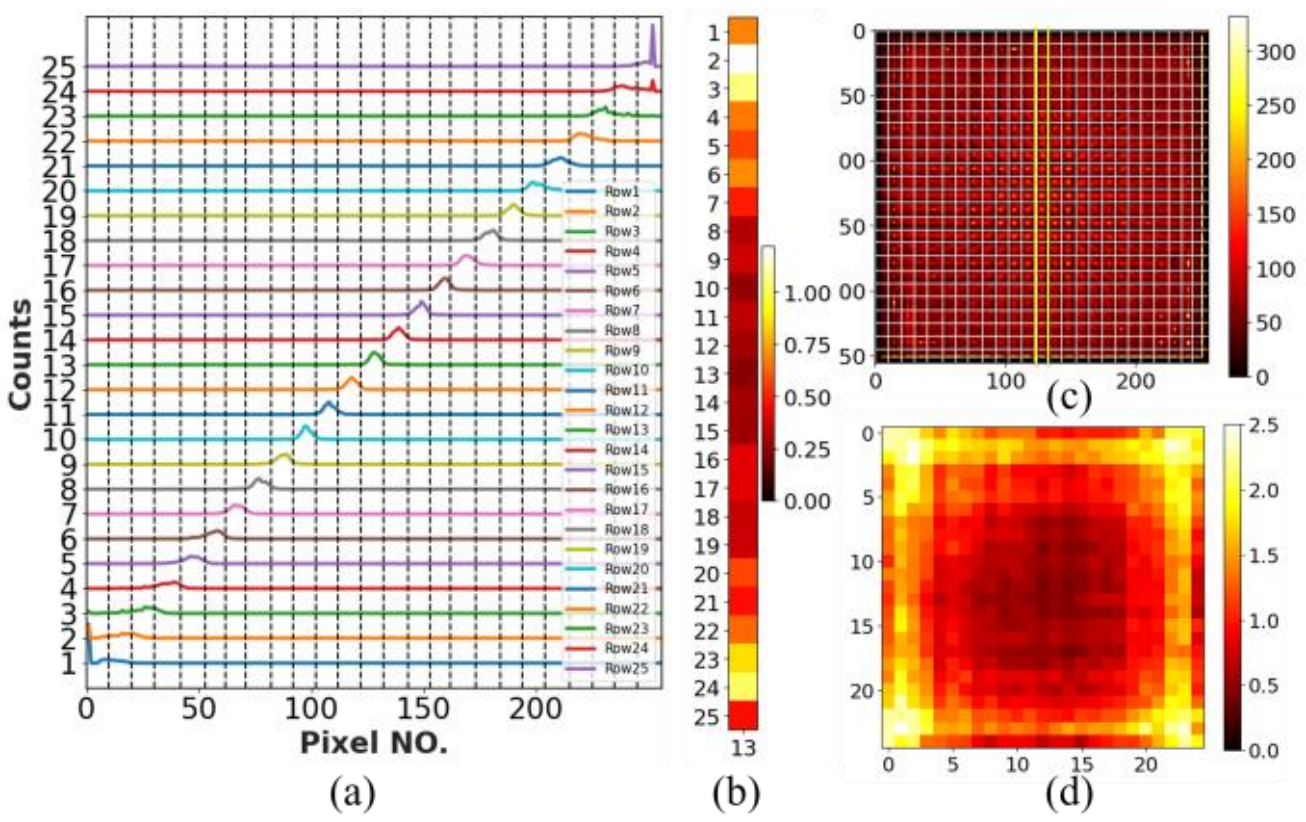

**Fig. 11.** Bilinear interpolation: (a) event-count profiles of the 25 collimation datasets along the central column; (b) positional errors of the selected datasets; (c) segmented decoding map, where the decoding region corresponding to the selected 25 datasets is highlighted between the two yellow lines; (d) positional errors of all 25 × 25 collimation datasets.

*Bicubic method:* Using the bicubic interpolation method, the decoding maps and event-count profiles of the collimated dataset were obtained, as shown in Fig. 12. The decoding maps demonstrated that this method was able to resolve nearly all 25 × 25 pixels with a relatively uniform distribution, which is advantageous for subsequent pixel segmentation. Even at the crystal edges and corners, several decoded spots remained clearly distinguishable. The mean PVRs, averaged over the row and column directions, were 3.36 and 3.63 for pixel resolutions of 256 and 512, respectively. Twenty-five distinct peaks corresponding to the 25 columns of pixels were clearly identified in the profiles. Although the edge crystals were successfully decoded, edge effects were still present, resulting in a noticeable decrease in the PVR near the boundaries.

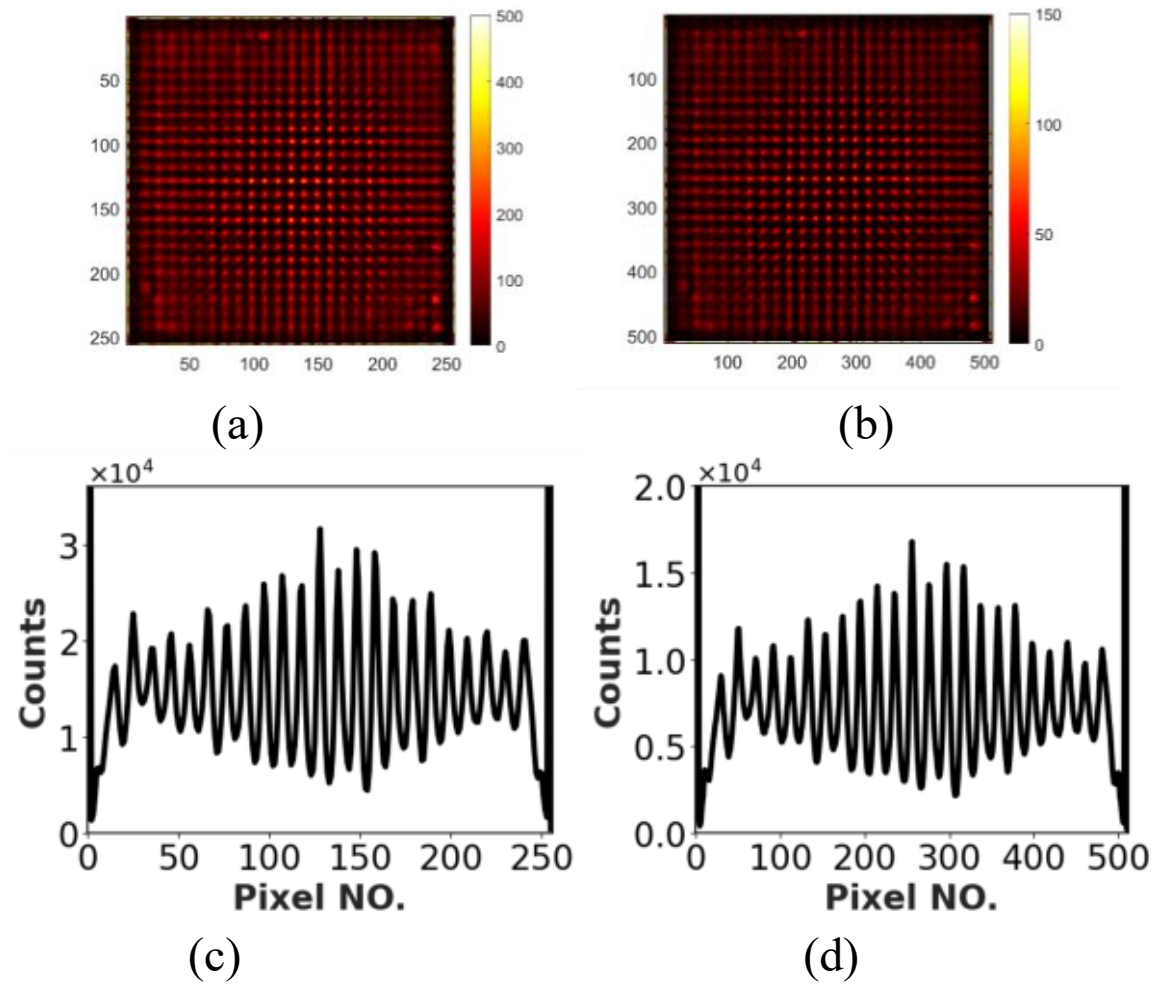

**Fig. 12.** Decoding maps of the collimated experiments obtained using bicubic interpolation with the ML algorithm at pixel resolutions of (a) 256 × 256 and (b) 512 × 512; corresponding event-count profiles are shown in (c) and (d), respectively.

Decoding error analysis was subsequently performed for the bicubic interpolation using the ML algorithm. The 25 collimation datasets along the central column were analyzed (Fig. 13(a) and (b)); the events were well localized in the central region, and the mean absolute error increased from 0.16 mm at the central pixels to 0.96 mm at the edges. Across all 25 × 25 pixels, the average positional error was 1.07 ± 0.45 mm (Fig. 13(d)). The fraction of counts decoded to the correct crystal was 58.9% overall (83.4% central 5 × 5, 48.9% outer ring), and 96.5% of the counts fell within ±1 crystal (98.0% central, 95.7% outer ring) (Fig. 16(c)).

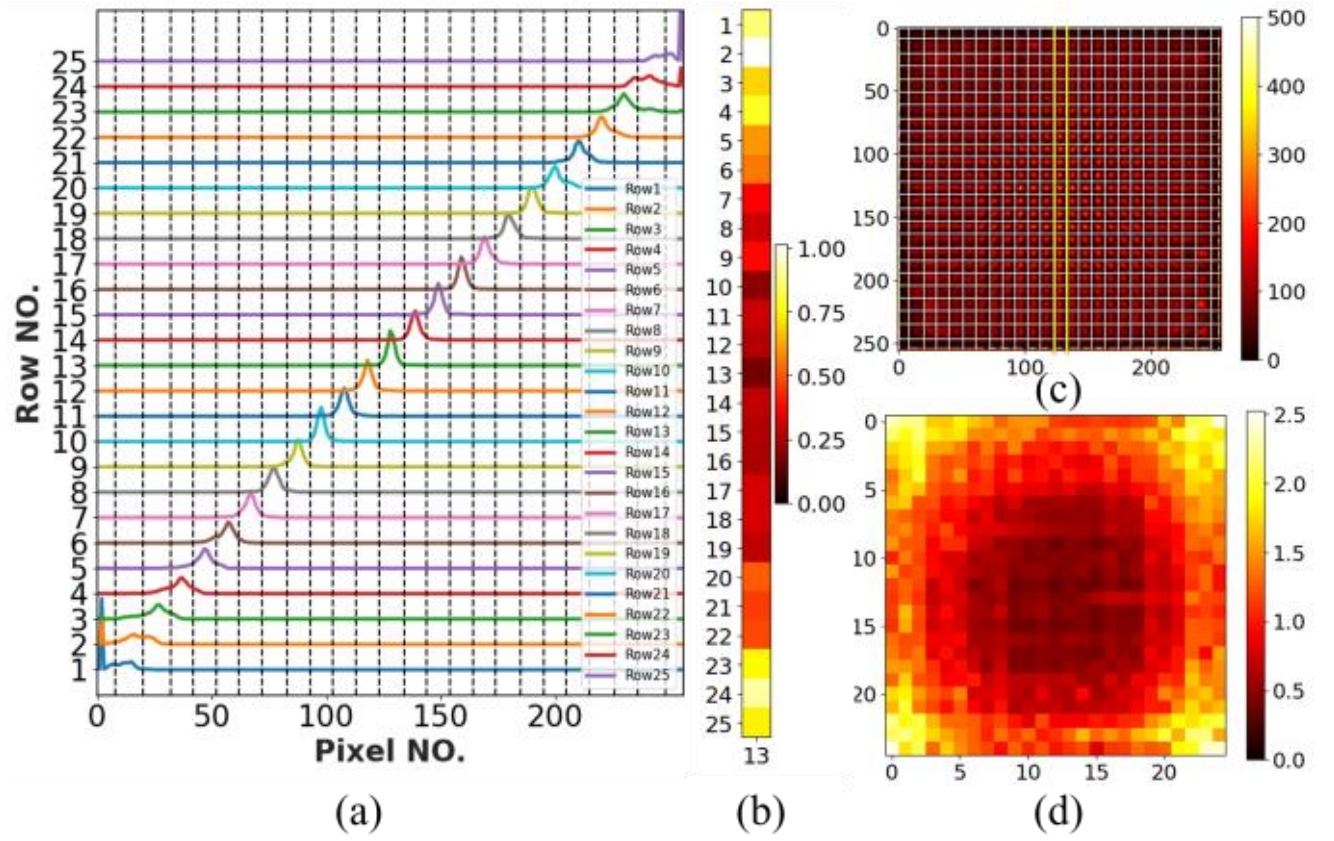

**Fig. 13.** Bicubic interpolation: (a) event-count profiles of the 25 collimation datasets along the central column; (b) positional errors of the selected datasets; (c) segmented decoding map, where the decoding region corresponding to the selected 25 datasets is highlighted between the two yellow lines; and (d) positional errors of all 25 × 25 collimation datasets.

*Nearest-neighbor method:* Fig. 14 shows the decoding results of the collimated dataset obtained using the nearest-neighbor interpolation at pixel resolutions of 256 × 256 and 512 × 512. Unlike the bilinear and bicubic methods, the nearest-neighbor approach produced discrete decoding results. In the decoding maps, all 25 × 25 decoding positions were clearly identified, and the corresponding event-count profiles exhibited 25 distinct peaks, indicating that the method successfully classified the collimated datasets into 25 × 25 discrete positions.

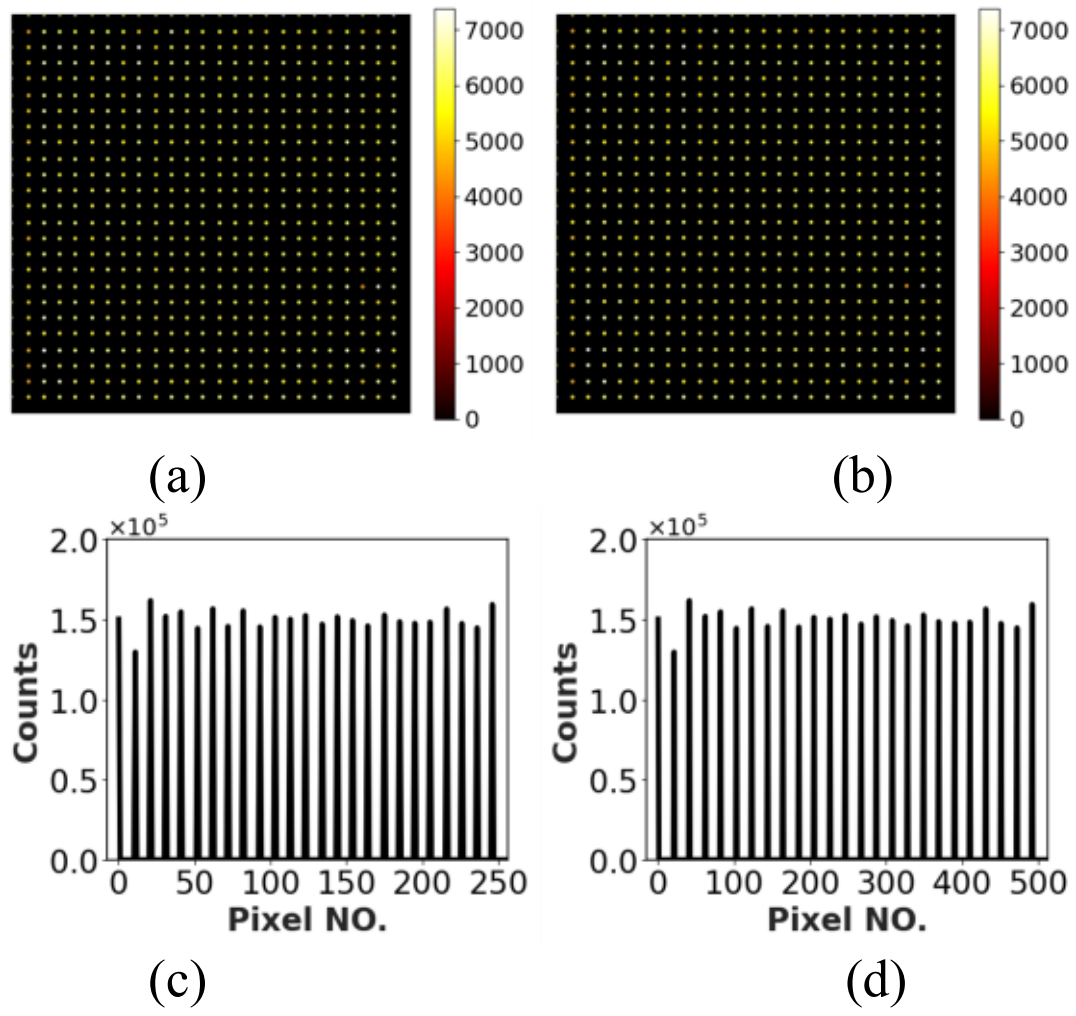

**Fig. 14.** Decoding maps of the collimated experiments obtained using nearest-neighbor interpolation with the ML algorithm at pixel resolutions of (a) 256 × 256 and (b) 512 × 512; corresponding event-count profiles are shown in (c) and (d), respectively. To enhance the contrast, the decoded pixels in (a) and (b) are each rendered as an individual marker, color-coded by the number of events assigned to that segment using a linear scale, so that the response of every segment is clearly visible.

For the nearest-neighbor interpolation, which produces discrete decoded positions, the 25 collimation datasets along the central column were analyzed (Fig. 15(a) and (b)); the mean absolute error ranged from 0.18 mm at the central pixels to 1.01 mm at the edges, and the average positional error across all 25 × 25 pixels was 1.00 ± 0.42 mm (Fig. 15(d)), the best of the four methods. It also yielded the highest fraction of counts decoded to the correct crystal, 61.3% overall (84.5% central 5 × 5, 52.2% outer ring), with 96.5% of the counts decoded within ±1 crystal (98.1% central, 95.6% outer ring) (Fig. 16(d)).

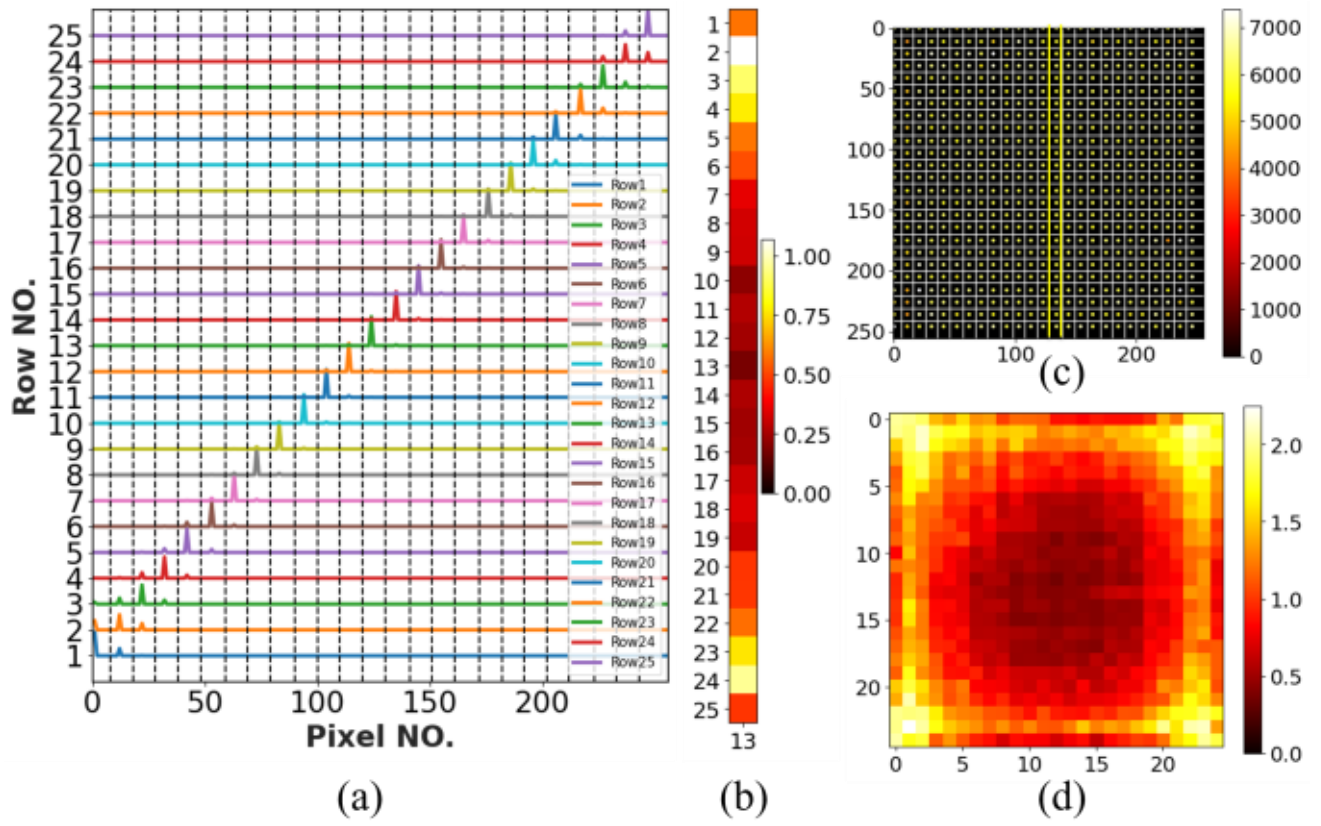

**Fig. 15.** Nearest-neighbor interpolation: (a) event-count profiles of the 25 collimation datasets along the central column; (b) positional errors of the selected datasets; (c) segmented decoding map, where the decoding region corresponding to the selected 25 datasets is highlighted between the two white lines; and (d) positional errors of all 25 × 25 collimation datasets.

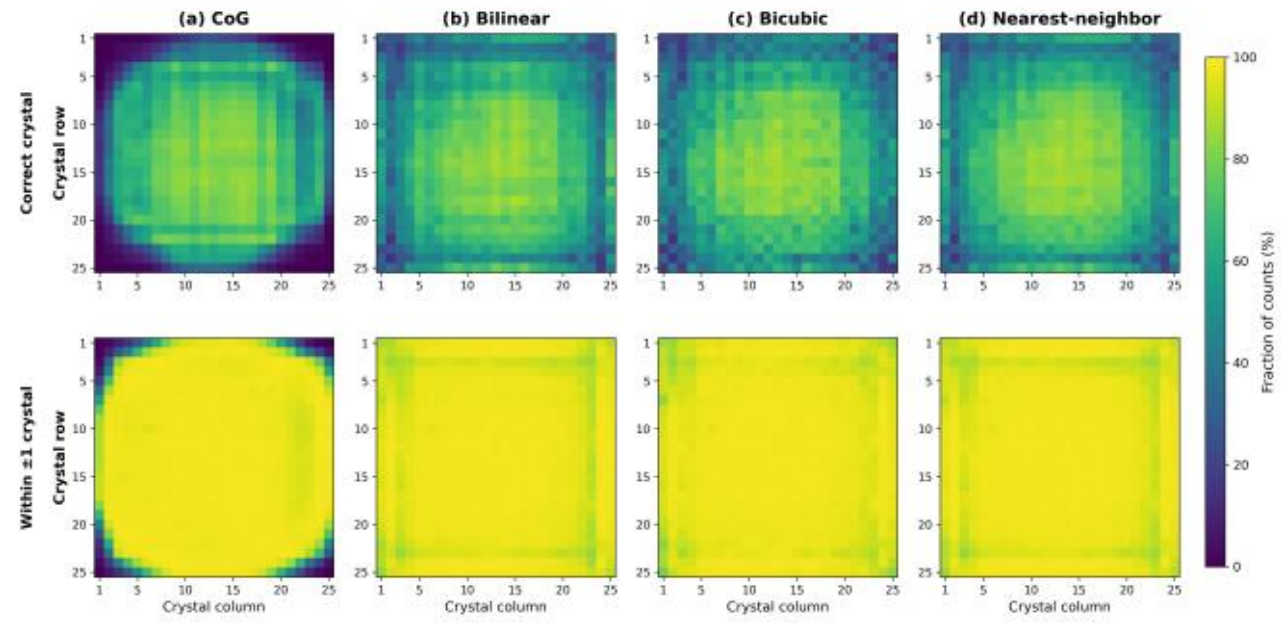

**Fig. 16.** Spatial distribution of the per-position fraction of counts decoded to the correct crystal (top row) and within ±1 crystal of the true position (bottom row) across the 25 × 25 crystal array, obtained using (a) the CoG method and the ML method with (b) bilinear, (c) bicubic, and (d) nearest-neighbor interpolation.

### 3) *Flood irradiation data*

Because the ML method demonstrated superior performance with the collimated dataset, its application was further extended to the flood-irradiation data. Fig. 17–19 present the decoding results obtained using bilinear, bicubic, and nearest-neighbor interpolation methods, respectively, at pixel resolutions of 256 and 512, together with the corresponding event-count profiles along the *x*-direction. Similar to the collimated results, the bilinear interpolation method exhibited limited discrimination among crystal pixels and introduced noticeable artifacts, as shown in Fig. 17. The mean PVRs, averaged over the row and column directions, were 3.84 and 5.40 for pixel resolutions of 256 and 512, respectively. Note, however, that these nominally high PVR values are inflated by the decoding artifacts—which concentrate counts into fewer, sharper spots—rather than reflecting genuine pixel discrimination, as evidenced by the fewer resolvable pixels and the inferior quantitative accuracy of the bilinear method in the collimated data.

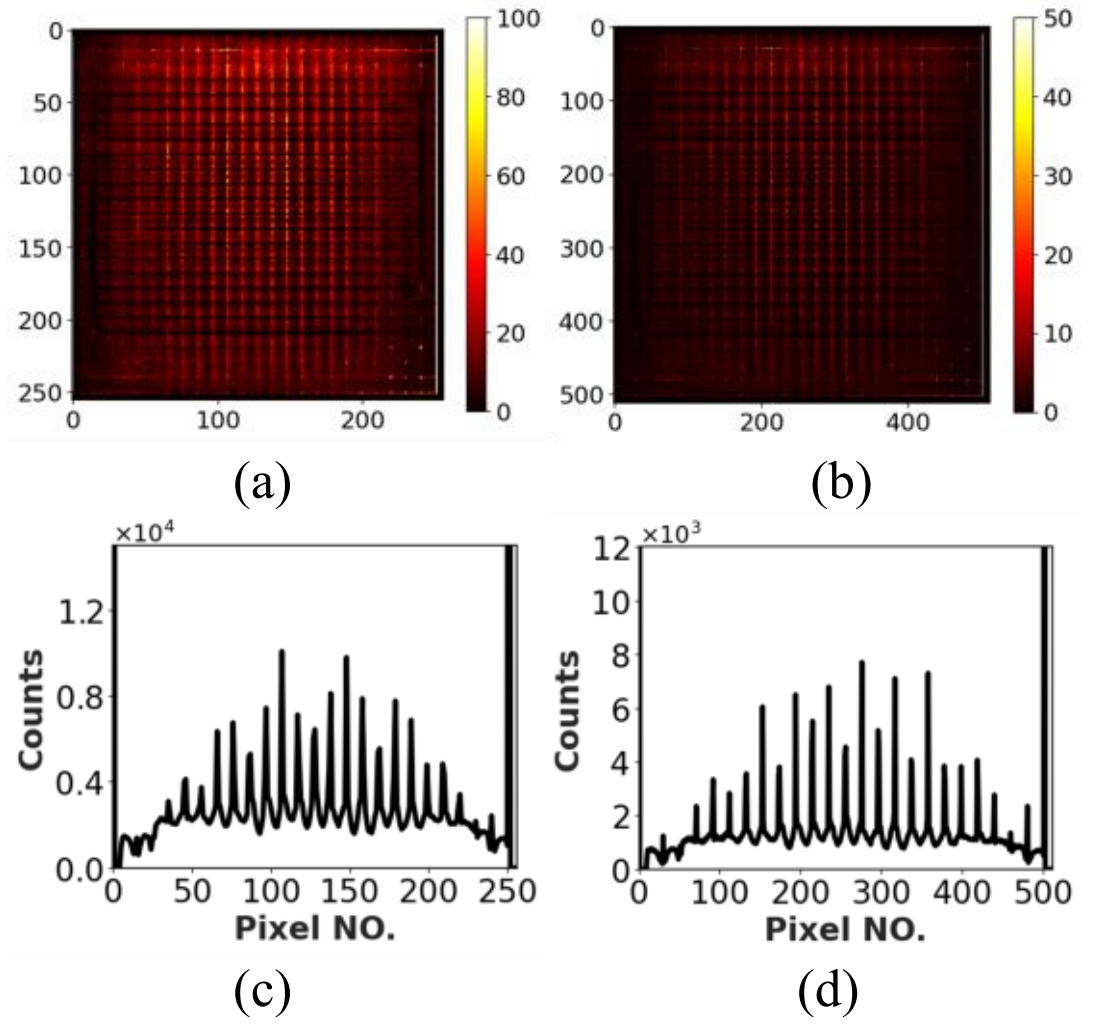

**Fig. 17.** Decoding maps of the flood-irradiation dataset obtained using bilinear interpolation with the ML algorithm at pixel resolutions of (a) 256 × 256 and (b) 512 × 512; corresponding event-count profiles are shown in (c) and (d), respectively.

The bicubic interpolation method successfully resolved most of the crystal pixels, with distinct pixel distributions observed even at the edges and corners, as shown in Fig. 18. The event-count profiles showed that although the decoding performance degraded toward the periphery, 25 peaks corresponding to the 25 rows of crystal pixel positions remained readily distinguishable. The mean PVRs, averaged over the row and column directions, were 2.96 and 3.29 for pixel resolutions of 256 and 512, respectively, enabling clear pixel discrimination and thereby improving positioning accuracy when compared with the CoG algorithm.

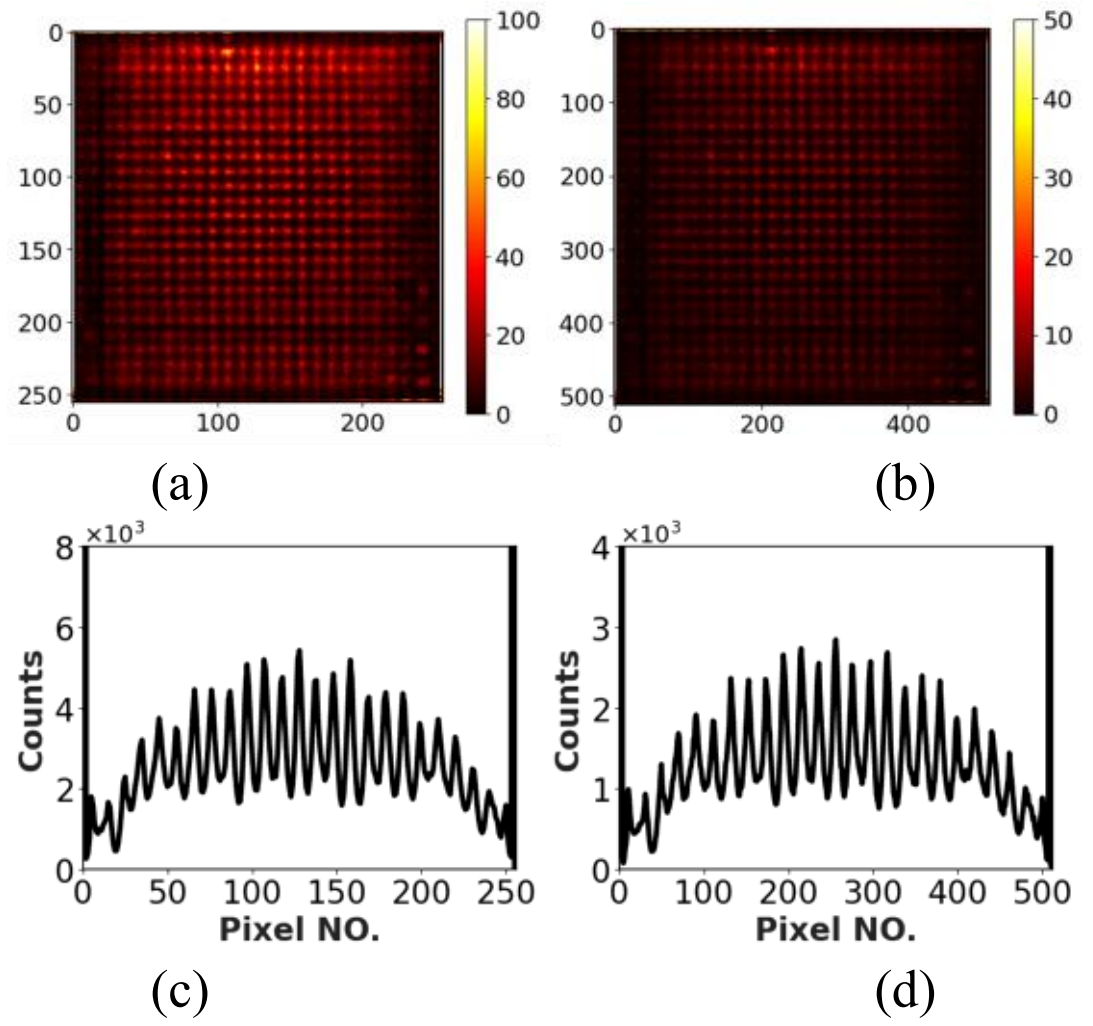

**Fig. 18.** Decoding maps of the flood-irradiation dataset obtained using bicubic interpolation with the ML algorithm at pixel resolutions of (a) 256 × 256 and (b) 512 × 512; corresponding event-count profiles are shown in (c) and (d), respectively.

The nearest-neighbor interpolation approach also achieved discrete decoding of the 25 × 25 positions, as shown in Fig. 19. Although its performance could not be evaluated using the PVR metric, the decoding results from the collimated dataset suggested that its positioning accuracy was slightly superior to that of the bicubic approach, with the added advantage of facilitating straightforward segmentation of the decoding maps.

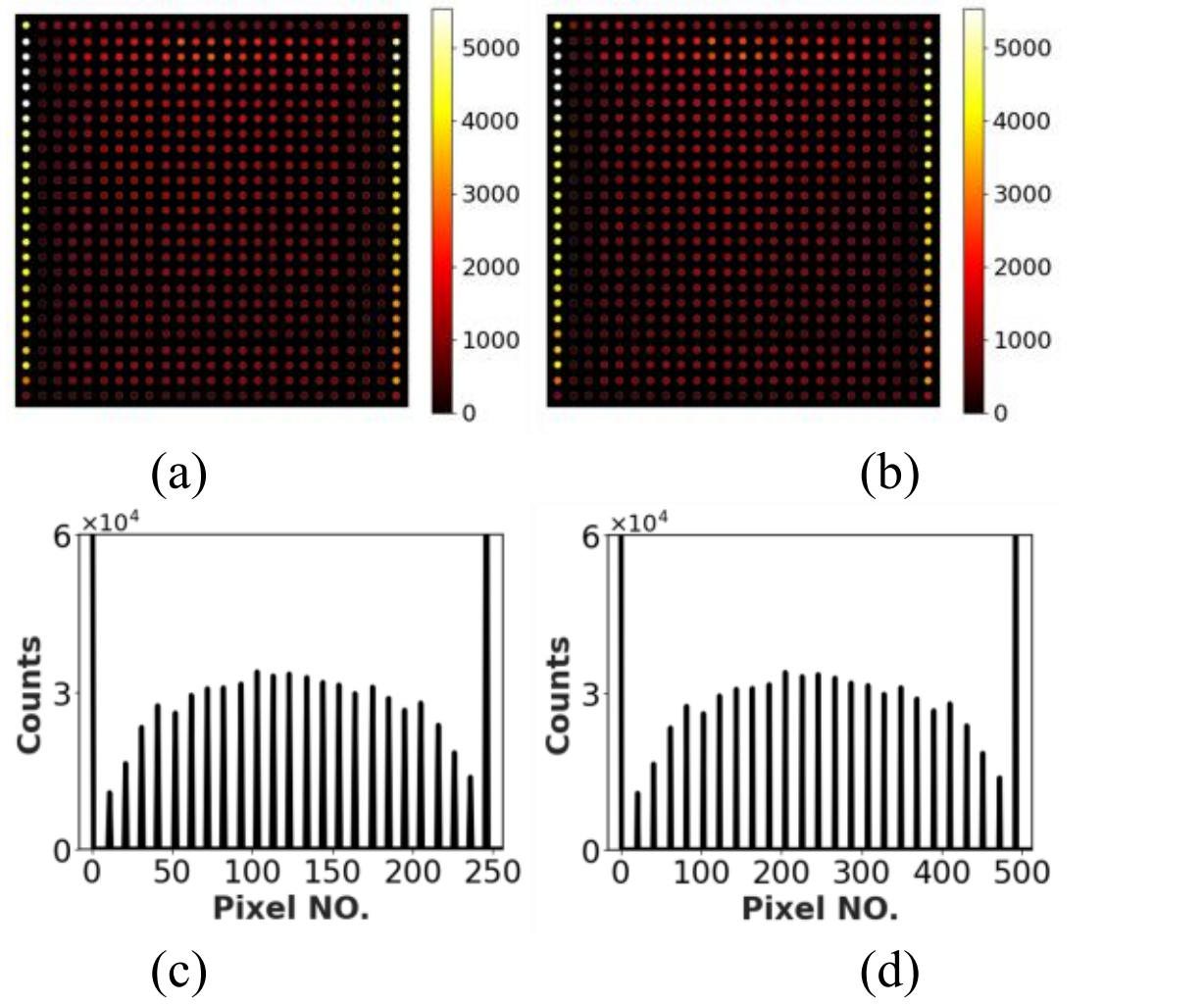

**Fig. 19.** Decoding maps of the flood-irradiation dataset obtained using nearest-neighbor interpolation with the ML algorithm at pixel resolutions of (a) 256 × 256 and (b) 512 × 512; corresponding event-count profiles are shown in (c) and (d), respectively. To enhance the contrast, the decoded pixels in (a) and (b) are each rendered as an individual marker, color-coded by the number of events assigned to that segment using a linear scale, so that the response of every segment is clearly visible.

Table I summarizes the decoding performance of the four methods. The ML-based methods outperformed the CoG algorithm on every quantitative metric; among them, the nearest-neighbor interpolation performed best, achieving the lowest positional error and the highest correct-crystal fraction. For all ML methods, more than 96% of the counts were decoded within ±1 crystal of the true position, compared with 90.3% for CoG, indicating that the residual assignment ambiguity is essentially confined to the immediately adjacent crystals. All methods exhibited an edge effect, with lower correct-crystal fractions in the outermost five-crystal ring than in the central 5 × 5 region; in the central region the four methods were nearly identical (within-±1 fraction 98.0–98.1%), and for the ML methods the outer-ring within-±1 fraction remained above 95%, whereas for CoG it dropped to 86.1%.

**TABLE I.** Summary of the decoding performance of the four positioning methods.

| Method | PVR, coll. (256/512) | PVR, flood (256/512) | Positional Error, mean ± std (mm) | Correct crystal fraction, Overall (central 5×5 / outer 5-ring) (%) | Within ±1 crystal fraction, Overall (central 5×5 / outer 5-ring) (%) |
|---|---|---|---|---|---|
| CoG | - | - | 1.44 ± 1.26 | 52.1 (81.0/38.8) | 90.3 (98.0/86.1) |
| Bilinear | 2.28/2.94 | 3.84/5.40 | 1.14 ± 0.48 | 57.3 (79.2/49.2) | 96.3 (98.0/95.5) |
| Bicubic | 3.36/3.63 | 2.96/3.29 | 1.07 ± 0.45 | 58.9 (83.4/48.9) | 96.5 (98.0/95.7) |
| Nearest-neighbor | * | * | 1.00 ± 0.42 | 61.3 (84.5/52.2) | 96.5 (98.1/95.6) |

- PVR could not be evaluated for the CoG method, which does not resolve enough distinguishable peaks

* PVR could not be calculated because the nearest-neighbor method produces discrete peaks, for which no valley is defined.

## IV. Discussion and conclusion

This study implemented a laser-processed optical barrier technique to fabricate converging pixels within a CsI:Tl crystal, aiming to enhance decoding performance while maintaining the inherent advantages of higher sensitivity and improved packing efficiency for SPECT detectors. The fabricated crystal featured converging pixels with a 2 mm pitch at the photodetector plane and a 1.6 mm pitch at the entrance surface. A custom-built four-axis motion platform was developed to enable high-precision calibration of the 25 × 25 converging pixels at specific incident angles. Both collimated and flood-irradiation datasets were decoded and compared using the CoG and ML methods. For the ML approach, three interpolation schemes—bilinear, bicubic, and nearest-neighbor—were investigated.

Beyond the decoding performance demonstrated above, the converging-pixel geometry confers a system-level advantage central to the impact of this work (Fig. 1). As introduced in Section I, aligning each pixel with the pinhole's line of collimation renders the detector response depth-invariant; its significance becomes clear in the context of depth-of-interaction (DOI) parallax, which fundamentally limits thick scintillation detectors in pinhole and converging-collimator SPECT. In parallel-pixel and monolithic crystals, obliquely incident photons interact at an unknown depth, and the resulting lateral mispositioning grows with crystal thickness and incidence angle; this couples detection efficiency to spatial resolution, so that the thicker crystals needed for high sensitivity inevitably blur the periphery of the field of view. Established remedies—multi-layer or phoswich detectors, dual-ended readout, and statistical DOI estimation in monolithic crystals—recover the missing depth information but add readout channels, dedicated electronics, or elaborate calibration[47]. The converging-pixel crystal instead removes the parallax geometrically in a single-layer, single-ended detector with no DOI decoding, making thick, high-sensitivity crystals usable without the customary resolution penalty—especially valuable for dedicated brain and small-animal SPECT, where high sensitivity and sub-2-mm resolution are required simultaneously. A quantitative, system-level evaluation of the resulting resolution–sensitivity trade-off, via Monte-Carlo simulation of a complete converging-pixel module, is the subject of ongoing work.

The ML method demonstrated strong capability in handling noise and incomplete data, exhibiting excellent positioning performance in the LIOB-processed CsI:Tl crystal. Compared with the CoG algorithm, the quality of the decoding map was significantly improved. Notably, in the edge and corner regions, where the CoG method failed to provide clear pixel separation, the ML method using both bicubic and nearest-neighbor interpolation achieved clear pixel separation. Beyond the PVR, the ML method also yielded markedly better quantitative accuracy than the CoG algorithm: the average positional error decreased from 1.44 ± 1.26 mm (CoG) to 1.00 ± 0.42 mm (nearest-neighbor), the fraction of counts decoded to the correct crystal increased from 52.1% to 61.3% overall, and the fraction decoded within ±1 crystal increased from 90.3% to 96.5%, with the largest improvement in the outermost five-crystal ring (within-±1 fraction from 86.1% to 95.6%). It should be noted that the CoG decoding map was likewise segmented into discrete pixel assignments using the same look-up-table procedure, so that all four methods were compared on an equal, discrete-pixel basis. The lower positional error and higher correct- and within-±1-crystal fractions of the ML methods therefore reflect genuinely higher decoding accuracy than CoG.

Among the three interpolation schemes, the bilinear method exhibited comparatively poorer decoding performance. For the complete collimated dataset, the overall decoding errors obtained with the bicubic and nearest-neighbor interpolation methods across the 25 × 25 positions were 1.07 ± 0.45 mm and 1.00 ± 0.42 mm, respectively. The superior performance of nearest-neighbor interpolation over bicubic may be attributed to the discrete, non-smooth nature of the detector response at pixel boundaries. The LIOB creates physical optical barriers between pixels, resulting in sharp discontinuities in the MDRF across pixel boundaries rather than smooth gradients. To our knowledge, this represents the first demonstration that statistical ML/MDRF decoding can completely resolve an LIOB optical-barrier, converging-pixel CsI:Tl crystal—a structure that the conventional CoG algorithm cannot fully decode—which we regard as a key methodological contribution of this work.

In the experiments, the collimated dataset was acquired using a $^{99m}$Tc source with a photopeak energy of 140 keV, whereas the flood-irradiation experiment employed a $^{57}$Co source with a photopeak energy of 122 keV. The flood-decoding maps indicated that variations in photon energy peaks had no significant influence on the applicability of the MDRF or the decoding performance of the ML method, as each converging pixel remained clearly distinguishable. These results suggest that the MDRF-based decoding is relatively robust to modest photopeak-energy differences, at least between 122 and 140 keV. Further measurements with additional radionuclides would be needed to establish broader energy independence.

In the CoG algorithm, applying an energy threshold or selecting the channel with the highest energy can effectively suppress electronic noise and enhance decoding performance. A similar strategy was applied in the ML decoding; however, it did not produce any significant improvement in the decoding results. This finding indirectly confirms the intrinsic capability of the ML method to handle noise and incomplete data. Furthermore, an energy-iterative ML decoding approach based on equation (5) was implemented, which likewise did not yield any measurable enhancement in performance. Several previous studies have attempted to optimize ML-based decoding performance through threshold analysis of likelihood probabilities. For example, in [48] the effects of applying a likelihood-based filter on system sensitivity, energy resolution, and image quality are investigated. Similarly, in [49] it is reported that comparing the maximized likelihood to a threshold is an effective way to reject events depositing energy across multiple pixels while retaining both photoelectric and Compton interactions. In the present study, a similar threshold-comparison approach was employed. For instance, the decoding probabilities obtained using the bicubic interpolation method for the collimated dataset are illustrated in Fig. 20. Here the decoding probability is the maximized log-likelihood, i.e., the value of $\mathcal{L}_i\left(i, E^{(n)}|\boldsymbol{q}\right)$ in equation (3) at the pixel i(n) selected by equation (4). The z-axis of Fig. 20 is therefore in log-likelihood units, and each pixel shows this quantity averaged over the events recorded at the corresponding collimation position. These values vary smoothly across the array, and the per-event likelihoods remain clustered. However, unlike in monolithic detectors, this optional likelihood-threshold filtering did not lead to a significant improvement in decoding performance. This is because the laser-induced optical barriers confine the scintillation light within each pixel, so that events depositing energy across multiple pixels occur only rarely and few events remain to be rejected. The absence of improvement therefore reflects an inherent advantage of the optical-barrier detector design rather than a limitation of the ML decoding method.

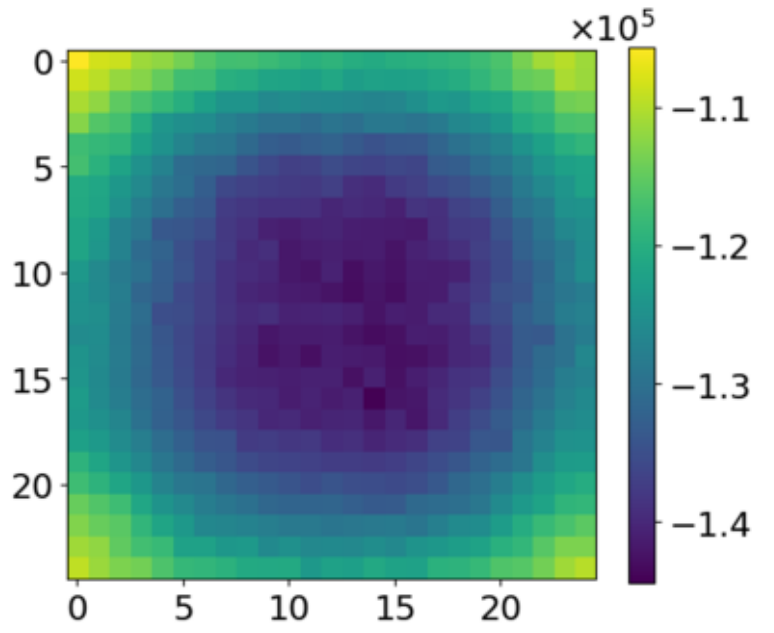

**Fig. 20.** Heat map of the average decoding probabilities obtained using the ML algorithm with bicubic interpolation for the collimated datasets.

In summary, this study demonstrated the feasibility of a laser-fabricated optical barrier technique for the design and fabrication of a converging-pixel CsI:Tl crystal. The results confirmed that the ML approach outperformed the CoG method, achieving clear and complete resolution of all crystal pixels. The successful fabrication and decoding of the proposed detector establish a solid foundation for the broader application of laser-processed scintillators in PET and SPECT systems, demonstrating their promising potential for high-resolution, high-sensitivity medical imaging. Future work will focus on employing artificial intelligence–based algorithms to further optimize the decoding process and to perform comparative analyses.

## Acknowledgment

We acknowledge the initial involvement of Salar Sajedi in readout electronics and Owen Anderson in the stage system. All authors declare that they have no known conflicts of interest in terms of competing financial interests or personal relationships that could have an influence or are relevant to the work reported in this paper.